\documentclass[amssymb,prb,aps,twocolumn,amsmath,showpacs]{revtex4}

\usepackage[dvips]{graphicx}
\usepackage{amstext}

\begin{document}

\title{Spin Triplet Supercurrent in Co/Ni Multilayer Josephson Junctions with Perpendicular Anisotropy}
\author{E. C. Gingrich, P. Quarterman, Y. Wang, R. Loloee, W. P. Pratt, Jr., and N. O. Birge}
\email{birge@pa.msu.edu} \affiliation{Department of Physics and
Astronomy, Michigan State University, East Lansing, Michigan
48824-2320, USA}

\date{\today}

\begin{abstract}

We have measured spin-triplet supercurrent in Josephson junctions of the form S/F'/F/F'/S, where S is superconducting Nb, F' is a thin Ni layer with in-plane magnetization, and F is a Ni/[Co/Ni]$_n$ multilayer with out-of-plane magnetization.  The supercurrent in these junctions decays very slowly with F-layer thickness, and is much larger than in similar junctions not containing the two F' layers.  Those two features are the characteristic signatures of spin-triplet supercurrent, which is maximized by the orthogonality of the magnetizations in the F and F' layers.  Magnetic measurements confirm the out-of-plane anisotropy of the Co/Ni multilayers.  These samples have their critical current optimized in the as-prepared state, which will be useful for future applications.

\end{abstract}

 \maketitle

Josephson junctions containing ferromagnetic (F) materials have received much attention in the past decade.\cite{BuzdinReview}  It was predicted 30 years ago that the supercurrent in such junctions should oscillate as a function of the F-layer thickness,\cite{Buzdin:82} but it wasn't until just over a decade ago that the oscillations were observed in experiments.\cite{Ryazanov:01,Kontos:02}  The oscillations arise because the two electrons from a spin-singlet Cooper pair in the superconductor (S) enter different spin bands in F, separated by the exchange energy.  In the diffusive limit, the pair correlations decay exponentially over a very short distance scale limited by the exchange energy in F.

Electron pairs in equal-spin spin-triplet states are not subject to the exchange energy, hence they are able to propagate long distances in F.  While superconductors with bulk spin-triplet pairing are rare, it was predicted in 2001 that spin-triplet pair correlations can be generated in S/F systems involving non-collinear magnetizations, even when all superconductors present in the system have traditional spin-singlet symmetry.\cite{Bergeret:01,Kadigrobov:01,Eschrig:03,Bergeret:05}  Some experimental evidence of the long range triplet appeared in 2006,\cite{Keizer:06,Sosnin:06} while more convincing evidence was released in 2010.\cite{Khaire:10,Robinson:10,Sprungmann:10,Anwar:10}  Our group's contribution\cite{Khaire:10} was based on the measurements of critical current, $I_c$, in Josephson junctions of the form S/F'/F/F'/S, where F was a synthetic antiferromagnet, and F' were thin ferromagnetic layers with magnetizations non-collinear with F.  In these experiments, F took the form of Co/Ru/Co, with Ru thickness of 0.6 nm.  The system possesses anti-parallel coupling of the two Co magnetizations resulting in zero net magnetic flux, which prevents distortion of the ``Fraunhofer patterns" one observes in plots of $I_c$ vs. $H$ applied parallel to the plane of the Josephson junction.  It was found that $I_c$ slowly decreased with Co thickness, as opposed to samples without the F' layers which showed a rapid decay in $I_c$.  The long range supercurrent in samples possessing F' layers was strong evidence for the presence of spin-triplet pair correlations.\cite{Houzet:07}

The next step was to attempt to optimize the samples so as to observe the maximum enhancement in $I_c$ for a given ferromagnetic thickness.\cite{Klose:12}  The F' layers were magnetized in the film plane at low temperatures.  During this process, the Co/Ru/Co layers also magnetize and undergo a ``spin-flop" transition, which results in the magnetization of the Co layers being in-plane and perpendicular to the F' layer magnetization after the field is removed.  The perpendicular magnetization configuration optimizes the spin-triplet generation, which is maximum for 90-degree non-collinearity of the ferromagnetic layers.\cite{Bergeret:03,Houzet:07}

The present work was motivated by a desire to find a suitable replacement for our Co/Ru/Co synthetic antiferromagnet so future work can take advantage of a virgin state magnetic configuration that is pre-optimized on fabrication.  An ideal central ferromagnetic system would naturally possess a magnetization perpendicular to the F' layers, and yet would not adversely effect the characteristic Fraunhofer patterns of such samples.  Such a material exists in the Co/Ni multilayer system.\cite{Daalderop:91}  When alternating layers of Ni and Co thin films are deposited, the system naturally aligns its magnetization perpendicular to the plane of the film.  This is due to the large magnetic anisotropy energy and the location of the Fermi energy near bands whose spin-orbit interaction favors a perpendicular anisotropy.\cite{Daalderop:91}

\begin{figure}[tbh!]
\begin{center}
\includegraphics[width=3.2in]{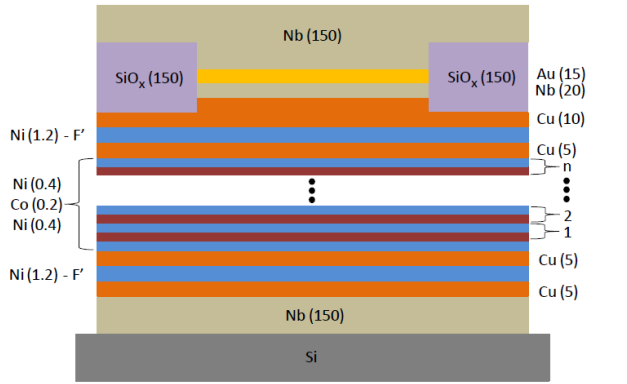}
\end{center}
\caption{(color online) Cartoon of our multilayer stack with F' layers. All thicknesses are in nm. The multilayer is sputtered in two steps, with the pillar defined by photolithography and ion milling. The Co/Ni multilayer in each sample has $n$ layers of Co and $n+1$ layers of Ni in the central ferromagnetic stack.}\label{Multilayer}
\end{figure}

In this work, we have grown two sets of samples with differing thin film ferromagnetic structures.  In the first set, we start by growing multilayers of the form Nb(150)/Cu(5)/Ni(0.4)/[Co(0.2)/Ni(0.4)]$_n$/Cu(10) /Nb(20)/Au(15), with all thicknesses in nm.  The second type of sample adds the F' layers in the form of two 1.2 nm layers of Ni inserted into the Cu spacers on either side of the Co/Ni multilayer structure, with all but the top most layer of Cu grown to a thickness of 5 nm as in Figure \ref{Multilayer}.  In both cases, the entire multilayer was formed without breaking vacuum.

\begin{figure}[tbh!]
\begin{center}
\includegraphics[width=3.2in]{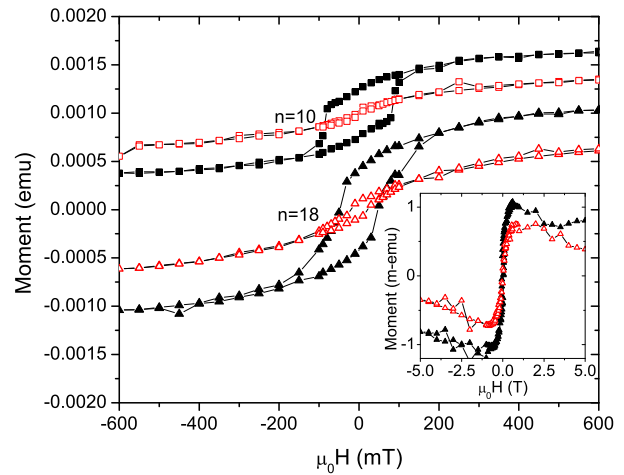}
\end{center}
\caption{(color online) Magnetometer measurements of samples with $n=10$ (squares) and $n=18$ (triangles) Co/Ni multilayers.  Black solid symbols represent data with the field applied perpendicular to the substrate, red open symbols are in the parallel orientation.  (Lines connecting the symbols are guides for the eye.)  Offset of 0.001 is added to the $n=10$ data for clarity.  Inset: Measurements (in units of 10$^{-3}$emu) of the $n=18$ sample up to 5 T indicate that the parallel orientation does not easily saturate to the same extent as does the perpendicular orientation. (The negative slope at high field is due to the diamagnetic contribution from the Si substrate.)}\label{MagData}
\end{figure}

To verify the perpendicular anisotropy of the Co/Ni multilayer structure, magnetization measurements were performed at 12 K in a commercial SQUID magnetometer.  Samples of the first multilayer type with varying $n$ were fabricated as a film with only 50 nm of Nb on the bottom, and 10 nm Nb on the top. These samples were measured with the field applied in both the in-plane (parallel) and out-of-plane (perpendicular) directions.  The data for the $n$ = 10 and $n$ = 18 samples are shown in Figure \ref{MagData}.  The perpendicular direction shows a stronger hysterisis than the parallel direction for all $n$, though the remnant magnetization is not near the saturation value in most samples.  At low fields the parallel configuration appears to saturate to half the value of the perpendicular direction.  The inset presents higher field measurements of the $n$ = 18 sample.  We attempted to magnetize the Co/Ni multilayer to high field in order to force the parallel orientation to reach the perpendicular saturation value.  While this did not occur at 5 T, the parallel configuration did more closely approach the perpendicular saturation.  This suggests that the perpendicular anisotropy is very strong, and restricts the complete canting of the moments into the plane.

Josephson Junctions with circular cross sections were formed using photolithography and ion milling as discussed in our previous work.\cite{Khaire:10,Klose:12,Khasawneh:09}  All data were taken on 10 $\mu$m-diameter pillars at 4.2K by dipping the sample into a liquid helium dewar.  The current-voltage (I-V) characteristics of the samples were measured using a current comparitor circuit using a SQUID as a null detector.  The samples exhibit the standard I-V characteristics for an overdamped Josephson junction,
\begin{equation}
V(I) = I/|I|*R_N \textrm{Re}[(I^2-I_c^2)^{1/2}]
\end{equation}
where $R_N$ is the normal-state resistance determined from the slope of the I-V relation at large currents.

\begin{figure}[tbh!]
\begin{center}
\includegraphics[width=3.2in]{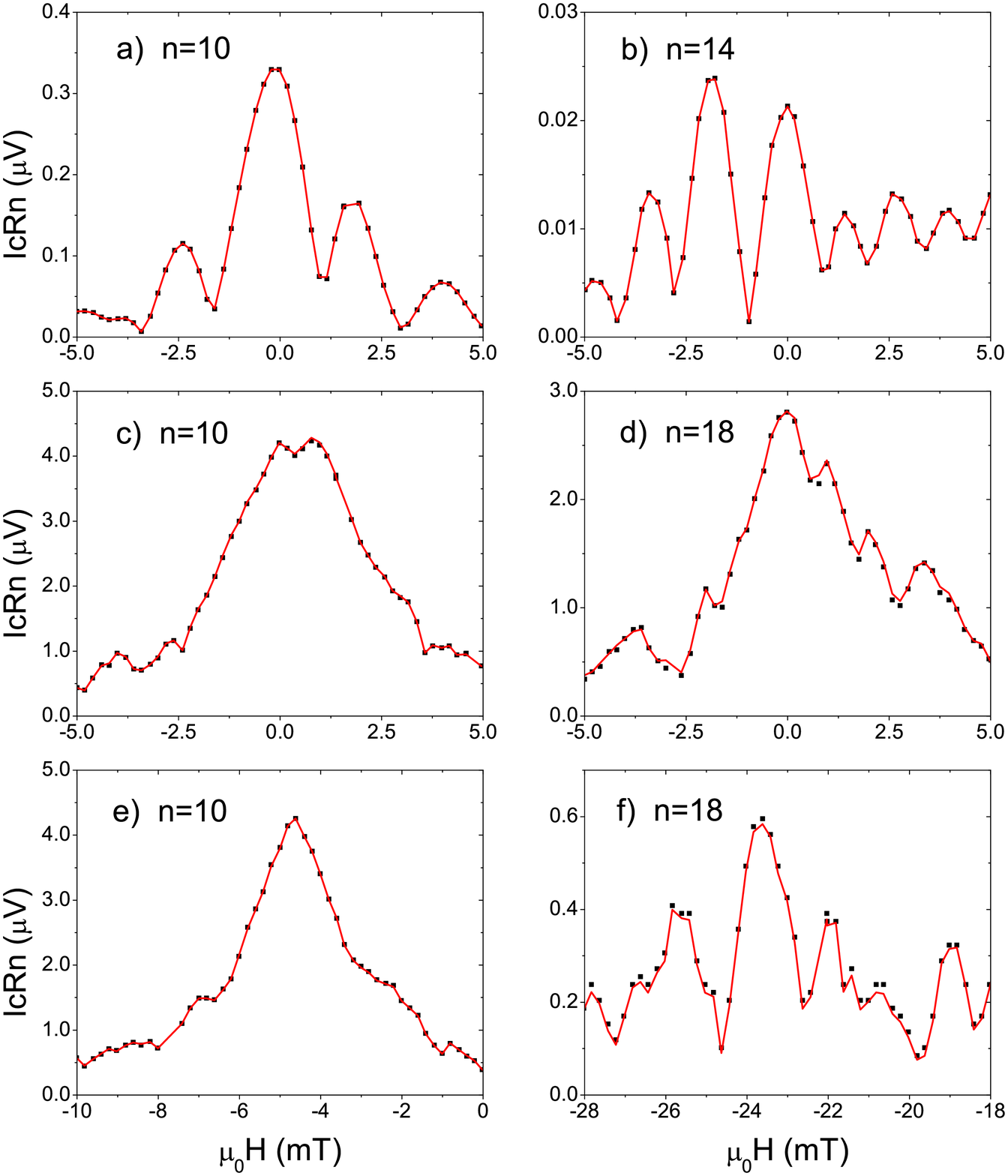}
\end{center}
\caption{(color online) Representative Fraunhofer patterns. All pillars are 10 $\mu$m in diameter. Lines are guides for the eye. Panels a) and b) show data for samples without F' layers, which carry only spin-singlet supercurrent.  Panels c) and d) show virgin-state data for samples with F' layers, carrying spin-triplet supercurrent, and panels e) and f) show data for the same two samples after they were exposed to a large in-plane field.  (Note that the central peak is shifted to negative field in the last two panels.) }\label{Fraunhofer}
\end{figure}

All of the samples are initially characterized by applying a small magnetic field in the plane of the substrate, perpendicular to the current direction.  Plotting $I_c$ vs. $H$ should yield the Fraunhofer pattern for our pillars (exactly, it should be the ``Airy pattern", since our pillars are circular).  Figure \ref{Fraunhofer} shows the $I_c$ vs. $H$ data for several of our 10 $\mu$m diameter junctions in the virgin state, both with and without F'.  Figure \ref{Fraunhofer}a is an example of a reasonably high quality Fraunhofer pattern, while Figure \ref{Fraunhofer}b demonstrates that not all of our pillars display pristine patterns.  This is due to the ferromagnetic materials present in the samples, which add internal magnetic fluxes that cause the Fraunhofer patterns to deviate from the ideal.\cite{Bourgeois:01,Khaire:09}  The samples including the F' layer tend to have more uniformly high quality patterns compared to those without F'.  This is not understood, but could be due to the addition of the first F' layer during fabrication, where the resulting fringing field may influence the domain structure of the Co/Ni layers during deposition. As with our previous samples containing Co/Ru/Co trilayers, the virgin state Fraunhofer patterns tend to fluctuate from run to run.\cite{Wang:12}

We attempted to magnetize the Co/Ni multilayer in the samples without F' in hopes of improving the Fraunhofer quality.  We magnetized the samples in the perpendicular direction in fields up to 0.4 T at room temperature (our dippers are currently arranged only for in-plane magnetic fields) and again characterized the $I_c$ vs. $H$ of the samples. There was no net enhancement of $I_c R_N$ over all samples. Though some samples did show a modest increase, just as many showed no change or even a decrease in peak $I_c$.  We also did not observe a notable improvement in the Fraunhofer quality over the measured samples.

\begin{figure}[tbh!]
\begin{center}
\includegraphics[width=2.8in]{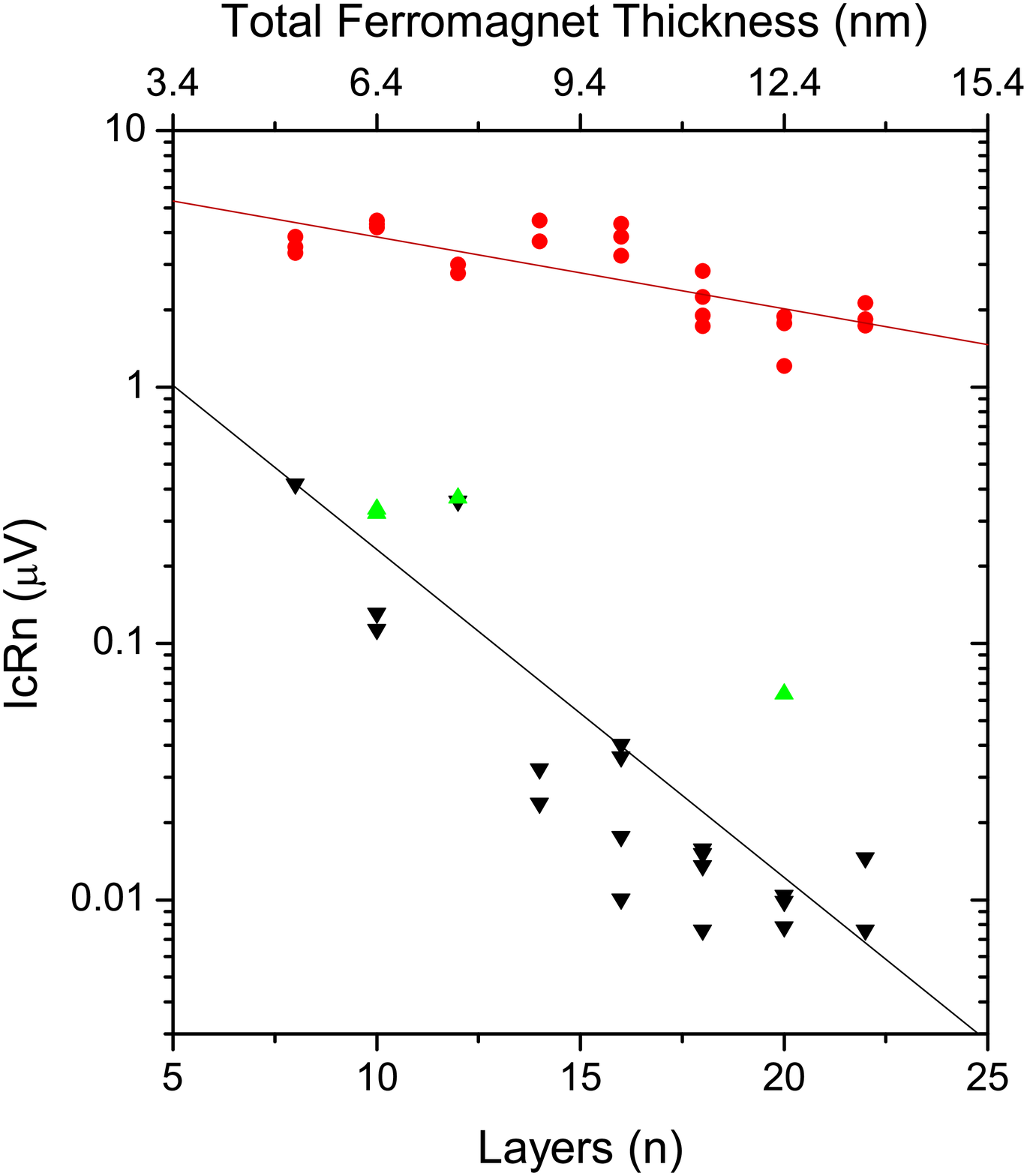}
\end{center}
\caption{(color online) This semi-log plot of $I_c R_N$ vs. number of layers shows a distinct enhancement in signal from the samples without F' (triangles) to the samples with F' (red circles). The slope of the decay of the samples without F' is approximately -0.29, compared to -0.06 for the samples with F'.  The upright (green) triangles represent the subset of samples without F' that gave the highest-quality Fraunhofer patterns.  The slope of the fit for those samples alone is -0.18.}\label{IcRn}
\end{figure}

Figure \ref{IcRn} presents our main result, a plot of $I_c R_N$ vs. the number $n$ of Co/Ni bilayers.  The black and green triangles represent the data for samples without the F' layers, while red circles are all samples possessing the F' layer.  There is a very strong enhancement in the $I_c R_N$ for those samples with an F' layer as compared to those without.  That enhancement is as much as a factor of 200 for the thickest samples with $n=22$.  In addition, the decrease of $I_c R_N$ with $n$ is much more gradual in the samples containing F' layers.  These two features are clear evidence of the generation of the spin triplet pairs in this system.

One can express the decay of $I_c R_N$ with $n$ as an exponential: $I_c R_N \propto exp[-A*n]$ with $exp[-A]$ the decay per Co/Ni bilayer.  From Figure \ref{IcRn}, we obtain $A=0.29\pm 0.04$ for the samples without F' (spin-singlet decay), and $A=0.064\pm 0.011$ for the samples with F' (spin-triplet decay).  (One can also re-write the relation as $I_c R_N \propto exp[-d_F/\xi_{F1}]$ with $\xi_{F}$ an effective decay length; then $A=0.6$nm$/\xi_{F}$.) To compare with previous work on S/F/S Josephson junctions, one can calculate the expected singlet decay in the samples without F' layers if one assumes that the Co/Ni multilayer acts as if all of the Ni films are combined into one, and similarly for Co.  Using singlet decay lengths from Khasawneh \textit{et al.}\cite{Khasawneh:11} for Co and Robinson \textit{et al.}\cite{Robinson:06} for Ni we calculate a value $A = 0.18$ for our system, which is somewhat less than our measured value of $A=0.29$.  This is not surprising, since the simple model used above does not account for any spin-memory loss at Ni/Co interfaces.\cite{Nguyen:10}  If one considers only the samples with the highest-quality Fraunhofer patterns (upright green triangles), the data suggest a value of $A=0.18\pm 0.05$, which happens to match the naive calculation neglecting interface effects.

In our previous work,\cite{Klose:12, Wang:12} we optimized the $I_c R_N$ products by exposing the samples to a large in-plane field of order 0.2-0.3 T while they were at 4.2 K.  That procedure had two effects: the Ni F' layers were magnetized in the direction of the field, while the Co layers in the Co/Ru/Co trilayer ended up with their in-plane magnetizations nearly perpendicular to the field, after undergoing a spin-flop transition.  We have carried out a similar procedure for the samples in this study containing F' layers, namely we exposed them to an in-plane field of 0.32 T at 4.2 K -- sufficient to completely magnetize the Ni layers.\cite{Klose:12}  Panels e) and f) in Figure \ref{Fraunhofer} show the results.  After magnetization, a shift in the Fraunhofer peaks was observed, as would be expected for the additional directed magnetic flux in the system. However, as $n$ increased the shift also grew, typically to 6 or 8 mT, but in one case to 24 mT (Figure \ref{Fraunhofer}f) -- much more than the 2.5 mT shift observed by our group with samples containing Co/Ru/Co and comparable Ni F' layers.\cite{Wang:12}  The larger field shift of the Fraunhofer patterns for the large $n$ samples suggests that the moments in the Co/Ni multilayers are canting somewhat toward the field direction.  That hypothesis is consistent with the in-plane magnetization data shown in Figure \ref{MagData}, which show a clear non-zero in-plane remnant magnetization in the $n=18$ sample, whereas the $n=10$ sample has almost no in-plane remnant magnetization. Upon being magnetized, the peak value of $I_c R_N$ stayed about the same in the samples with $n=8$ and 10, but decreased in the samples with higher $n$, again with the $n=18$ sample shown in Figure \ref{Fraunhofer}f being the extreme case where $I_c R_N$ dropped by a factor of 5.

To compare the absolute magnitude of $I_c R_N$ to our previous results with samples containing Co/Ru/Co,\cite{Klose:12,Wang:12} it is important to note that the previous samples in the virgin state show signals much smaller than they do when magnetized. Depending on sample to sample conditions, this can be as much as a factor of 20 change in signal.  Ultimately, the optimized values of $I_c R_N$ for the previous samples peak at about 1.5 $\mu$V for samples with a total Co thickness of 12 nm.\cite{Wang:12}  For a similar thickness of the Co/Ni multilayer in the new samples, Figure \ref{IcRn} shows comparable peak values in $I_c R_N$ for the virgin state data.

In conclusion, we have measured the supercurrent of S/F/S Josephson junctions containing Co/Ni multilayers as the middle layer.  There is a strong enhancement in the $I_c R_N$ product of junctions containing additional F' layers on both sides of the central F layer, indicating that spin-triplet supercurrent is generated in those samples.  In the virgin state, the samples demonstrate an equal or larger $I_c R_N$ than when magnetized.  The virgin-state values of $I_c R_N$ are comparable to the optimized magnetized values of the previous generation of samples containing Co/Ru/Co.  This is consistent with the known perpendicular anisotropy of the Co/Ni multilayer system, with magnetization oriented 90 degrees with respect to the in-plane magnetizations of the Ni F' layers.  Thus, the Co/Ni multilayer is an excellent replacement for the Co/Ru/Co and provides a promising system for future experiments in spin-triplet physics.

Acknowledgements: We acknowledge the helpful contributions of I. Beskin that enabled improvements to our measuring system.  We also thank B. Bi for technical assistance, and the use of the W.M. Keck Microfabrication Facility.  Research supported by the U.S. Department of Energy, Office of Basic Energy Sciences, Division of Materials Sciences and Engineering under Award DEFG02-06ER46341.


\begin{thebibliography} {99}

\bibitem{BuzdinReview} A.I. Buzdin, Rev. Mod. Phys. {\textbf{77}} 935 (2005).

\bibitem{Buzdin:82} A. Buzdin, L.N. Bulaevskii, and S.V. Panyukov, JETP Lett. \textbf{35}, 178
(1982).

\bibitem{Ryazanov:01} V.V. Ryazanov, V.A. Oboznov, A.Yu. Rusanov, A.V. Veretennikov, A.A. Golubov, and J. Aarts, Phys. Rev. Lett. \textbf{86}, 2427 (2001).

\bibitem{Kontos:02} T. Kontos, M. Aprili, J. Lesueur, F. Genet, B. Stephanidis, and R. Boursier, Phys. Rev. Lett. \textbf{89}, 137007 (2002).

\bibitem{Bergeret:01} F.S. Bergeret, A.F. Volkov, and K.B. Efetov, Phys. Rev. Lett. \textbf{86}, 4096 (2001).

\bibitem{Kadigrobov:01} A. Kadigrobov, R.I. Shekhter, and M.
Jonson, Europhys. Lett. \textbf{54}(3), 394 (2001).

\bibitem{Eschrig:03} M. Eschrig, J. Kopu, J.C. Cuevas, and G. Sch\"{o}n, Phys. Rev. Lett., 90, 137003 (2003).

\bibitem{Bergeret:05} F.S. Bergeret, A.F. Volkov, K.B. Efetov, Rev. Mod. Phys. \textbf{77}, 1321 (2005).

\bibitem{Keizer:06} R.S. Keizer, S.T.B. Goennenwein, T.M. Klapwijk,
G. Miao, G. Xiao, and A. Gupta, Nature (London) \textbf{439}, 825
(2006).

\bibitem{Sosnin:06} I. Sosnin, H. Cho, V.T. Petrashov, and A.F.
Volkov, Phys. Rev. Lett. \textbf{96}, 157002 (2006).

\bibitem{Khaire:10} T.S. Khaire, M.A. Khasawneh, W.P. Pratt Jr. and N.O. Birge, Phys. Rev. Lett. \textbf{104},
137002 (2010).

\bibitem{Robinson:10} J.W.A. Robinson, J.D.S. Witt and M.G. Blamire, Science \textbf{329}, 59 (2010).

\bibitem{Sprungmann:10} D. Sprungmann, K. Westerholt, H. Zabel, M. Weides and H.
Kohlstedt, Phys. Rev. B \textbf{82}, 060505 (2010).

\bibitem{Anwar:10} M. S. Anwar, F. Czeschka, M. Hesselberth, M. Porcu, and J. Aarts, Phys. Rev. B \textbf{82}, 100501
(2010).

\bibitem{Houzet:07} M. Houzet and A.I. Buzdin, Phys. Rev. B \textbf{76}, 060504(R) (2007).

\bibitem{Klose:12} C. Klose, T.S. Khaire, Y. Wang, W.P. Pratt, Jr., N.O.
Birge, B.J. McMorran, T.P. Ginley, J.A. Borchers, B.J. Kirby, B.B.
Maranville, and J. Unguris, Phys. Rev. Lett. \textbf{108}, 127002
(2012).

\bibitem{Bergeret:03} A.F. Volkov, F.S. Bergeret, and K.B. Efetov,  Phys. Rev. Lett. \textbf{90}, 117006 (2003).

\bibitem{Daalderop:91} G.H.O. Daalderop, P.J. Kelly, and F.J.A. den Broeder,
Phys. Rev. Lett. \textbf{68}, 682 (1992).

\bibitem{Khasawneh:09} M.A. Khasawneh, W.P. Pratt Jr. and N.O. Birge,
Phys. Rev. B \textbf{80}, 020506(R) (2009).

\bibitem{Bourgeois:01} O. Bourgeois, P. Gandit, J. Lesueur, A.
Sulpice, X. Grison, and J. Chaussy, Eur. Phys. J. B \textbf{21},
75 (2001).

\bibitem{Khaire:09} T.S. Khaire, W.P. Pratt, Jr., and N.O. Birge,
Phys. Rev. B \textbf{79}, 094523 (2009).

\bibitem{Wang:12} Y. Wang, W.P. Pratt, Jr., and N.O. Birge,
Phys. Rev. B \textbf{85}, 214522 (2012).

\bibitem{Khasawneh:11} M.A. Khasawneh, T.S. Khaire, C. Klose, W.P. Pratt Jr. and N.O.
Birge, Supercond. Sci. Technol. \textbf{24}, 024005 (2011).

\bibitem{Robinson:06} J.W.A. Robinson, S. Piano, G. Burnell, C. Bell, and M.G. Blamire, Phys. Rev. Lett. \textbf{97}, 177003 (2006).

\bibitem{Nguyen:10} H.Y.T. Nguyen, R. Acharyya, E. Huey, B. Richard, R. Loloee, W.P. Pratt, Jr., J. Bass, S. Wang, and K. Xia, Phys. Rev. B \textbf{82}, 220401(R) (2010) measure spin-memory loss in Co/Ni multilayers with much thicker Co and Ni layers.

\end{thebibliography}
\end{document}